\documentclass[journal=jctcce,manuscript=article]{achemso}

\usepackage[version=3]{mhchem} 

\usepackage{amsmath,amssymb}
\usepackage{graphicx,color,subfigure}
\usepackage{amsfonts}
\usepackage{comment}
\usepackage{placeins}
\usepackage{multirow}

\usepackage{balance}
\usepackage{xspace}
\usepackage{algorithm,algorithmic}

\usepackage{appendix}

\usepackage{epsfig,amssymb,amsmath,amsthm,amsfonts,amsbsy,mathrsfs}
\usepackage{graphicx,color}
\usepackage{amsmath}
\usepackage{amssymb}
\usepackage{epstopdf}

\usepackage{lastpage}
\usepackage[format=plain,justification=raggedright,singlelinecheck=false,font=small,labelfont=bf,labelsep=space]{caption}

\allowdisplaybreaks

\usepackage[version=3]{mhchem} 

\newcommand{\vt}{\vec{\theta}}

\definecolor{MatlabY}{rgb}{0.929,0.694,0.125}

\author{Jie Liu}
\affiliation{Hefei National Laboratory for Physical Sciences at the
Microscale, University of Science and Technology of China, Hefei, Anhui 230026,
China}

\author{Zhenyu Li}
\email{zyli@ustc.edu.cn (Zhenyu Li)}
\affiliation{Hefei National Laboratory for Physical Sciences at
the Microscale, University of Science and Technology of China,
Hefei, Anhui 230026, China}

\author{Jinlong Yang}
\email{jlyang@ustc.edu.cn (Jinlong Yang)}
\affiliation{Hefei National Laboratory for Physical Sciences at the
Microscale, Department of Chemical Physics, and Synergetic
Innovation Center of Quantum Information and Quantum Physics,
University of Science and Technology of China, Hefei, Anhui 230026,
China}

\title{An efficient adaptive variational quantum solver of the Schr\"odinger equation based on reduced density matrices}

\begin{document}
\begin{abstract}
Recently, an adaptive variational algorithm termed Adaptive Derivative-Assembled Pseudo-Trotter ansatz Variational Quantum Eigensolver (ADAPT-VQE) has been proposed by Grimsley et al. (Nat. Commun. 10, 3007) while the number of measurements required to perform this algorithm scales $\mathcal{O}(N^8)$. In this work, we present an efficient adaptive variational quantum solver of the Schr\"odinger equation based on ADAPT-VQE together with the reduced density matrix reconstruction approach, which reduces the number of measurements from $\mathcal{O}(N^8)$ to $\mathcal{O}(N^4)$. This new algorithm is quite suitable for quantum simulations of chemical systems on near-term noisy intermediate-scale hardware due to low circuit complexity and reduced measurement. Numerical benchmark calculations for small molecules demonstrate that this new algorithm provides an accurate description of the ground-state potential energy curves. In addition, we generalize this new algorithm for excited states with the variational quantum deflation approach and achieve the same accuracy as ground-state simulations.
\end{abstract}

\maketitle
\section{Introduction}
The electronic structure problem is one of the most appealing applications of quantum computing~\cite{AspDutLov05,DuXuPen10,PerMcCSha14,MalBabKiv16,KanMezTem17,HemMaiRom18,AruAryBab20,Pre18,CaoRomOls19,McAEndAsp20}. To simulate electronic structure properties of molecules and solids, quantum algorithms, such as quantum phase estimation (QPE)~\cite{AspDutLov05,WhiBiaAsp11} and variational quantum eigensolver (VQE) algorithms~\cite{PerMcCSha14,McCRomBab16,RomBabMcC18}, had been proposed for solving the Schr\"odinger equation on a quantum computer. Quantum algorithms offer an exponential speedup of simulating many-electron systems over classical methods. The QPE algorithm evolves in time a quantum state under the molecular Hamiltonian of interest, which often requires circuits with a long gate sequence even for few qubits. By contrast, the VQE requires a much short coherence time at the cost of a large number of quantum measurements. Therefore, the VQE is a promising algorithm for chemistry simulations on near-term noisy intermediate-scale hardware with massive parallelization~\cite{MalBabKiv16}. 

The VQE algorithm applies the Rayleigh-Ritz variational principle to optimize the parameterized wave function, which minimizes the energy functional at convergence. The variational optimization procedure is often referred as a hybrid quantum-classical strategy, in which the calculations of various physical properties, such as the energy and gradient, are performed on the quantum computer and the update of parameters is performed on the classical computer~\cite{McCRomBab16,McCKimCar17,RomBabMcC18,ColRamDah18}. In the VQE framework, the state preparation and measurement should be repeated many times in order to measure the expectation value of operators. Therefore, to establish an advanced VQE algorithm, it is critical to reduce the cost of state preparation and measurement. 

The state preparation is closely related to the wave function ansatz. Unitary coupled cluster (UCC) is one of the most widely used wave function ansatzes in the VQE algorithm~\cite{PerMcCSha14,SheZhaZha17,RomBabMcC18,LeeHugHea19}. The UCC wave function with a finite cluster expansion can be easily prepared on a quantum computer. Recently, UCC with single and double excitations (UCCSD) has been successfully applied in experimental and theoretical chemistry simulations within the VQE framework~\cite{PerMcCSha14,SheZhaZha17}. UCCSD is more robust than traditional counterpart due to the fact that the unitary cluster operator involves not only excitation operators but also de-excitation operator. Nevertheless, UCCSD also suffers from the well-known problem of the single reference method when treating strongly correlated systems. 

Given a many-electron system described by the Hamiltonian containing up to two-body terms, it has been suggested that the ground-state wave function can be highly accurately represented by an exponential cluster expansion involving general two-body operators~\cite{Noo00,PieKowFan03,Dav03,Ron03,GriEcoBar19}. Thus, UCCGSD that extends UCCSD to include general one- and two-body operators~\cite{Noo00} has been introduced in the experimental demonstration of VQE on a photonic quantum processor~\cite{PerMcCSha14}. The classical numerical simulations reveal that the UCCGSD wave function is far more robust and accurate than the simpler UCCSD wave functions for some challenging problems, such as dissociation of $N_2$~\cite{LeeHugHea19}. However, one main drawback of UCCSD and UCCGSD is the requirement of a finite-order Suzuki-Trotter decomposition of the cluster operator in order to be implemented on a quantum computer~\cite{PouHasWec15,BabMcCWec15}. This leads to the so-called "Trotterization" error and the nonuniqueness of the operator ordering that has a significant effect on strongly correlated systems. An alternative scheme recently proposed by Grimsley {\it et al.}, termed Adaptive Derivative-Assembled Pseudo-Trotter (ADAPT) ansatz~\cite{GriEcoBar19}, aims at generating a maximally compact sequence of unitary transformations. ADAPT is appealing due to the minimum gate depth while the number of measurements for the residual gradients (see Equation~\eqref{eq:acse_res} in Section~\ref{sec:ADAPT}) used to determine the operator sequence scales $\mathcal{O}(N^8)$, where $N$ is the number of spin orbitals.

In this work, we propose an efficient adaptive variational quantum algorithm based on reduced density matrices (RDM). First of all, we reformulate the residual gradients with the two-electron RDM (2-RDM) and three-electron RDM (3-RDM), which explicitly reduces the scaling of measurements in ADAPT to $\mathcal{O}(N^6)$. Furthermore, we propose an approximate ADAPT ansatz termed ADAPT-V, in which the 3-RDM is computed as a functional of the 2-RDM using Valdemoro's reconstruction~\cite{ColPerVal93,ColVal93,ColVal94}. In the context of quantum computing, the ADAPT-V ansatz has two key attractive aspects: i) It generates a compact wave function, which promises low gate depth; ii) It requires only the 2-RDM in the variational optimization procedure, which means that the residual gradients are computed without any additional measurements. However, the ADAPT-V ansatz must update many operators at a time due to the approximation introduced in the residual gradients in order to guarantee the accuracy of the converged wave function. Alternatively, we propose another ansatz termed ADAPT-Vx, which first builds an auxiliary operator pool using Valdemoro's reconstruction of 3-RDM and then determines operators to be updated with the exact residual gradients. The ADAPT-Vx ansatz is able to generate a wave function as compact as the ADAPT ansatz while it requires slightly increasing measurements compared with the ADAPT-V ansatz. In addition, the ADAPT-V and ADAPT-Vx ansatzes can be straightforwardly generalized to excited-state simulations with the variational quantum deflation (VQD) algorithm based on the orthogonal constraint~\cite{HigWanBri19,JonEndMcA19}. 

This paper is organized as follow. Section~\ref{sec:theory} gives a brief description of the theoretical methodology, covering the ADAPT ansatz, the ADAPT-V and ADAPT-Vx ansatz reformulated with the reduced density matrices, the ADAPT-V and ADAPT-Vx VQE algorithms and their extension for excited states. In section~\ref{sec:resource}, we give a short discussion of quantum resource requirement for different wave function ansatzes. In section~\ref{sec:results}, we compute the ground-state and excited-state potential energy curve of $H_6$ and $N_2$ and analyze errors for different wave function ansatzes. A summary and outlook is given in Section~\ref{sec:conclusion}.

\section{Theory}\label{sec:theory}
The Hamiltonian of a many-electron system is expressed in the second-quantized as
\begin{equation}\label{eq:Hamiltonian}
\begin{split}
    \hat{H} &= \sum_{pq} h^p_q a^\dag_p a_q + \frac{1}{2} \sum_{pqrs} v^{pq}_{sr}  a^\dag_p a^\dag_q a_r a_s \\
            &= \frac{1}{2}\sum_{pqrs} H^{pq}_{sr}  a^\dag_p a^\dag_q a_r a_s,
\end{split}
\end{equation}
where $h^p_{q}$ are the one-electron integrals, including kinetic energy and ionic potential, and $v^{pq}_{rs}$ are the two-electron integrals. The reduced Hamiltonian matrix elements are written as
\begin{equation}
    H^{pq}_{rs} = \frac{1}{N-1}(h^p_r \delta^q_s + \delta^p_r h^q_s) +  v^{pq}_{rs},
\end{equation}
where $N$ is the number of electrons. $a_p^\dag$ and $a_p$ are the second-quantized creation and annihilation operators which satisfy the anticommutation relations. Here and after, only general spin molecular orbitals are involved in the definition of the wave function ansatz. 

The ground state wave function is obtained by solving
\begin{equation}
    \hat{H} |\Psi\rangle = E |\Psi \rangle. 
\end{equation}
The energy can be written as a functional of the 2-RDM,
\begin{equation}
\begin{split}
    E &= \langle \Psi | \hat{H} | \Psi \rangle \\
    &=\sum_{pqrs} H^{pq}_{rs} \, {}^2 D^{pq}_{sr}.
    \end{split}
\end{equation}
Here, the $m$-RDM is defined as
\begin{equation}
    ^m D^{p_1,\ldots,p_m}_{q_1,\ldots,q_m} = \frac{1}{m!} \langle \Psi| a_{p_1}^\dag \cdots a_{p_m}^\dag a_{q_m} \cdots a_{q_1} |\Psi \rangle.
\end{equation}

\subsection{Adaptive Derivative-Assembled Pseudo-Trotter ansatz}\label{sec:ADAPT}
Recently, Grimsley et al.~\cite{GriEcoBar19} suggested that the exact wave function can be represented as an arbitrarily long product of general one- and two-body exponentiated operators,
\begin{equation}\label{eq:wave0}
    | \Psi (\vec{\theta}) \rangle = \prod_k^{N_k} \prod_u^{N_{op}} e^{\theta_{u,k} \tau_u} |\Psi_0 \rangle,
\end{equation}
where $\tau_u \in \{\tau^p_q,\tau^{pq}_{rs}\}$ defines the anti-Hermitian operator pool $\mathcal{O}$
\begin{equation}
    \begin{split}
        &\tau^p_q=a_p^\dag a_q - a_q^\dag a_p \\
        &\tau^{pq}_{rs} = a_p^\dag a_q^\dag a_r a_s - a_s^\dag a_r^\dag a_q a_p.
    \end{split}
\end{equation}
$N_{op}$ is the number of operators in $\mathcal{O}$ and $N_k$ is the number of layers of the product. The reference state $|\Psi_0\rangle$ can be either a single-reference state, such as Hartree-Fock wave function, or a multireference state. 

In order to generate a maximally compact sequence of operators at convergence, the ADAPT wave function is iteratively updated by restricting $N_{op}$ to be 1 in Eq.~\eqref{eq:wave0}, which is simply written as
\begin{equation}\label{eq:wave}
    | \Psi (k) \rangle =  e^{\theta_k \tau_k} |\Psi(k-1) \rangle
\end{equation}
where $|\Psi(0)\rangle = |\Psi_0\rangle$ is the reference state. The energy functional in the $k$-th iteration is minimized with respect to parameters $\vec{\theta}$ by
\begin{equation}\label{eq:energy}
    E(k) = \min_{\{\theta_l\}_{l=1}^{k}} \{ \langle \Psi(k) | \hat{H} | \Psi(k) \rangle \}.
\end{equation}
The analytical gradient of the energy functional is formulated as
\begin{equation}\label{eq:acse_grad}
\begin{split}
    G_{l} &= \frac{\partial E(k)}{\partial \theta_l} \\
    &= \langle \Psi(k) | \hat{H} \prod_{m=l+1}^{k} e^{\theta_m \tau_m} \tau_l \prod_{n=1}^l e^{\theta_n \tau_n} |\Psi_0 \rangle  + c.c. 
    \end{split}
\end{equation}

Given a wave function in the form of Eq.~\eqref{eq:wave}, its accuracy can be assessed through gradients 
\begin{equation}\label{eq:acse_res}
\begin{split}
    R_u &= G_{k}|_{\theta_{k}=0,\tau_{k}=\tau_u} \\
    &= \langle \Psi(k-1)| [\hat{H},\tau_u] |\Psi(k-1) \rangle,
    \end{split}
\end{equation}
which are also referred as the residual errors of anti-Hermitian contracted Schr\"odinger equation (ACSE)~\cite{Maz06a,Maz07}. In order to distinguish them from gradients of Eq.~\eqref{eq:acse_grad} used in a gradient-based minimization algorithm, we call Eq.~\eqref{eq:acse_res} as the residual gradients. It is natural to choose the operator with the largest residual gradient to update the wave function. The convergence criteria can be defined as
\begin{equation}\label{eq:acse_conv}
    |\vec{R}|_2 = \sqrt{\sum_u |R_u|^2} < \epsilon.
\end{equation}
As a consequence, the ADAPT ansatz also represents an exact solution of the ACSE. Compared with UCC, ADAPT generates a well-defined and maximally compact ordering of the operators, which avoids errors from Trotterization and generates a low-depth quantum circuit. 

\subsection{Reduced density matrices}
In the ADAPT ansatz with general one- and two-body operators, both the number of operators in the operator pool and terms in the Hamiltonian scale as $\mathcal{O}(N^4)$. Therefore, the number of terms one should measure for the explicit calculations of the residual gradients with Eq.~\eqref{eq:acse_res} scales as $\mathcal{O}(N^8)$. On near-term noisy quantum hardware with limited size and number of quantum processor, it is necessary to reduce the scaling of measurements for practical simulations of chemical systems using the ADAPT ansatz. Here, we propose to reformulate the residual gradients with reduced density matrices in order to realise measurement reduction.

For one-body operators, Eq.~\eqref{eq:acse_res} can be rewritten in terms of 1- and 2-RDM's,
\begin{equation}\label{eq:acse_grad_approx_1b}
    \begin{split}
        &\langle \Psi|[a_{p_1}^\dag a_{q_1},\hat{H}] | \Psi \rangle \\
        =&\sum_{q_2}h^{q_1}_{q_2} \, ^1 D^{p_1}_{q_2} - \sum_{p_2} h^{p_2}_{p_1} \, ^1 D^{p_2}_{q_1} \\
        +& 2\sum_{p_3,q_2,q_3} v^{q_1p_3}_{q_3q_2} \, ^2 D^{p_1p_3}_{q_3q_2} - 2\sum_{p_2,p_3,q_3} v^{p_2p_3}_{p_1q_3} \, ^2 D^{p_2p_3}_{q_1q_3}.
    \end{split}
\end{equation}
For two-body operators, Eq.~\eqref{eq:acse_res} can be rewritten in terms of 2- and 3-RDM's,
\begin{equation}\label{eq:acse_grad_approx_2b}
    \begin{split}
        &\langle \Psi|[a_{p_1}^\dag a_{p_2}^\dag a_{q_1} a_{q_2},\hat{H}] | \Psi \rangle \\
        =& 2 \sum_{q_3} h^{q_2}_{q_3} {^2D}^{p_1,p_2}_{q_3,q_1} - 2\sum_{q_3} h^{q_1}_{q_3} {^2 D}^{p_1,p_2}_{q_3,q_2} + 2 \sum_{q_3,q_4} v^{q_2q_1}_{q_4q_3} {^2D}^{p_1,p_2}_{q_4,q_3}\\-&2 \sum_{q_3} h^{q_3}_{p_1} {^2 D}^{q_3,p_2}_{q_2,q_1} + 2\sum_{q_3} h^{q_3}_{p_2} {^2 D}^{q_3,p_1}_{q_2,q_1}
         + 2\sum_{p_3,p_4} v^{p_4p_3}_{p_1p_2} {^2D}^{p_3,p_4}_{q_2,q_1} \\ -& 6\sum_{p_3,q_3,q_4} v^{q_2p_3}_{q_4q_3} {^3D}^{p_1,p_2,p_3}_{q_4,q_3,q_1} 
        +  6\sum_{p_3,q_3,q_4} v^{q_1p_3}_{q_4q_3} {^3D}^{p_1,p_2,p_3}_{q_4,q_3,q_2} \\
    -& 6 \sum_{p_3,p_4,q_3} v^{p_4p_3}_{p_1q_3} {^3D}^{p_3,p_4,p_2}_{q_2,q_1,q_3} 
   + 6 \sum_{p_3,p_4,q_3} v^{p_3p_4}_{q_3,p_2} {^3D}^{p_3,p_4,p_1}_{q_2,q_1,q_3}.
    \end{split}
\end{equation}
It is clear that the 4-RDM is no more explicitly required to compute the residual gradients. The ADAPT ansatz using Eq.~\eqref{eq:acse_grad_approx_1b} and Eq.~\eqref{eq:acse_grad_approx_2b} is referred as ADAPT-RDM in this work. To perform the ADAPT-RDM VQE algorithm exactly, the calculations of the residual gradients in terms of 2- and 3-RDM's scale only as $\mathcal{O}(N^6)$, a significant reduction of measurements compared with the original ADAPT ansatz. However, the ADAPT-RDM is still too expensive for quantum simulations of large chemical systems on near-term quantum hardware with limited resources.

In order to avoid the explicit requirement of the 3-RDM in ADAPT, an alternative approach is to reconstruct the 3-RDM from the 2-RDM by the cumulant theory~\cite{Maz98,Maz98b,Maz00,Maz04,KutMuk99,KutMuk04}
\begin{equation}\label{eq:3-RDM}
\begin{split}
    ^3D_{q_1q_2q_3}^{p_1p_2p_3}&={^1D}_{q_1}^{p_1} \wedge {^1D}_{q_2}^{p_2} \wedge {^1D}_{q_3}^{p_3} \\
    &+ 3 \, {^2 \Delta}_{q_1q_2}^{p_1p_2} \wedge {^1D}_{q_3}^{p_1} + {^3 \Delta}_{q_1q_2q_3}^{p_1p_2p_3}.
    \end{split}
\end{equation}
Here
\begin{equation}
    {^2 \Delta}_{q_1q_2}^{p_1p_2} = {^2D}_{q_1q_2}^{p_1p_2} - {^1D}_{q_1}^{p_1} \wedge {^1D}_{q_2}^{p_2},
\end{equation}
and
\begin{equation}
        {^1D}_{q_1}^{p_1} \wedge {^1D}_{q_2}^{p_2} 
        =\frac{1}{2}({^1D}_{q_1}^{p_1}  {^1D}_{q_2}^{p_2} - {^1D}_{q_1}^{p_2} {^1D}_{q_2}^{p_1}),
\end{equation}
The operator $\wedge$ denotes the antisymmetric tensor product, known as the Grassmann wedge product. Neglecting the cumulant 3-RDM,
\begin{equation}
    {^3 \Delta}_{qst}^{ijk} = 0,
\end{equation}
which yields a first-order reconstruction of the 3-RDM from the 1- and 2-RDM's, termed Valdemoro's  reconstruction~\cite{ColPerVal93,ColVal93,ColVal94}. Here, we refer the ADAPT-RDM ansatz using Valdemoro's reconstruction as ADAPT-V. Hence, the calculations of the residual gradients require only 1- and 2-RDM's in the ADAPT-V.

\subsection{VQE algorithm}
In variational quantum algorithms, the wave function is often represented as a unitary transformation acting on a reference state,~\cite{McCRomBab16}
\begin{equation}\label{eq:unitary_transformation}
    |\Psi(\vec{\theta})\rangle = U(\vec{\theta})|\Psi_0\rangle.
\end{equation}
$U(\vec{\theta})$ can be either a single unitary operator or an finite long product of a series of unitary operators. The energy and wave function are determined through Rayleigh-Ritz variational principle,
\begin{equation}\label{eq:RR}
    E_0 = \min_{\vt}  \langle \Psi(\vec{\theta}) | \hat{H} | \Psi(\vec{\theta}) \rangle.
\end{equation}

In the ADAPT VQE algorithm, one key aspect is to grow the ansatz with the operator that is expected to restore the greatest amount of the correlation energy. In the ADAPT-V VQE algorithm, the approximate residual gradients is also expected to identify such an operator when the wave function is far away from the exact wave function. However, as the convergence is approaching, the relative deviations in the residual gradients becomes significant. Consequently, the ADAPT-V VQE algorithm may converge very slowly due to the inaccurate identification of the operator. In order to overcome this problem, an alternative solution is to increase the number of operators to be updated in each iteration. Instead of Eq.~\eqref{eq:wave}, the wave function in the form of Eq.~\eqref{eq:wave0} is adopted in this work
\begin{equation}\label{eq:waveV}
    | \Psi (\vec{\theta}) \rangle = \prod_k^{N_k} \prod_{u \in M} e^{\theta_{u,k} \tau_u} |\Psi_0 \rangle
\end{equation}
where $M$ contains $N_u$ operators with the largest residual gradients. The ordering of $N_u$ operators may be still different from the exact one while the influence of approximate residual gradients is suppressed as long as these ''important'' operators have been included in $M$.

The ADAPT-V VQE algorithm is summarized in Algorithm~\ref{alg:adapt}. In contrast with the original ADAPT VQE algorithm, the main difference of this new algorithm is presented as follows:
\begin{itemize}
    \item In Step 5, we compute $\{\vec{R}\}$ with Eq.~\eqref{eq:acse_grad_approx_1b} and ~\eqref{eq:acse_grad_approx_2b}, in which the 3-RDM is reconstructed from the 2-RDM using Valdemoro's reconstruction.   
    \item In Step 6, $N_u$ operators with largest absolute residual gradients are appended to the operator sequence $\{\vec{\tau}\}$. 
\end{itemize}

Given that Valdemoro's reconstruction can give a good estimation of the residual gradients, we propose an ADAPT-Vx ansatz to avoid a large $N_u$ required in the ADAPT-V. For ADAPT-Vx, Step 6 in Algorithm~\ref{alg:adapt} is modified as
\begin{itemize}
    \item $N_m$ operators with largest absolute residual gradients form an auxiliary operator pool $M'$. 
    \item The residual gradients for operators in $M'$ are recalculated with Eq.~\eqref{eq:acse_res}. 
    \item $N_u$ operators with largest absolute residual gradients in $M'$ are appended to the operator sequence $\{\vec{\tau}\}$. 
\end{itemize}
The number of operators in the auxiliary operator pool $M'$ is often chosen to be much less than $N_{op}$ in Eq.~\eqref{eq:wave0}. While $N_u$ operators refined from $M'$ are expected to have the maximum overlap with corresponding operators identified in the ADAPT ansatz. Therefore, ADAPT-Vx promises a more stable convergence as shown later at the cost of slightly increasing measurements scaling as $N_mN^4$.

\begin{algorithm}[H]
\caption{The ADAPT-V VQE algorithm for optimizing the wave function and the energy.}

\leftline{\textbf{Input:} Reference state $|\Psi_0\rangle$ and Hamiltonian $\hat{H}$.}

\leftline{\textbf{Output:} The energy and wave function of the target state.}

\begin{algorithmic}[1]

\STATE Prepare the initial wave function $|\Psi \rangle = |\Psi_0 \rangle$ in qubit representation.

\STATE Define the operator pool $\mathcal{O}$.

\STATE Initialize the operator sequence $\vec{\tau}=\{\}$ and parameters $\vec{\theta}=\{0\}$.

\WHILE {$|\vec{R}|_2 > \epsilon$}

\STATE Compute \{$\vec{R}$\} with Eq.~\eqref{eq:acse_grad_approx_1b} and Eq.~\eqref{eq:acse_grad_approx_2b} using Valdemoro's reconstruction for all $\tau_u \in \mathcal{O}$.

\STATE $\vec{\tau} \gets \{\vec{\tau}, \tau_{1},\ldots,\tau_{N_u}\} $ where $\{\tau_l\}|_{l=1}^{N_u} $ are $N_u$ operators with the largest absolute residual gradients and $\vec{\theta}=\{\vec{\theta},0\}$.

\STATE Update the new wave function with Eq.\eqref{eq:waveV} and the new energy functional with Eq.~\eqref{eq:RR}.

\STATE Optimize parameters $\vec{\theta}$.

\ENDWHILE

\STATE Return E($\vec{\theta}$) and $|\Psi(\vec{\theta})\rangle$.

\end{algorithmic}
\label{alg:adapt}
\end{algorithm}


\subsection{Excited-state approach}
The calculation of excited-state properties is essential to explore the photophysical and photochemical processes in molecules\cite{DreHea05} and solid state materials\cite{KimGor02}. Due to the recent advent in experimental and theoretical simulations of molecular ground-state properties on quantum computers, there is increasing interest in exploring excited states within the framework of VQE. Recent development of VQE algorithms for excited states includes the quantum subspace expansion (QSE) algorithm inspired by the linear response approach~\cite{McCKimCar17}, the folded spectrum (FS) algorithm by finding the expectation of the squared Hamiltonian~\cite{WanZun94,PerMcCSha14}, the witness-assisted variational eigenspectra solver (WAVES) algorithm that minimises the von Neumann entropy~\cite{SanWanGen18}, and the variational quantum deflation (VQD) algorithm based on the orthogonal constraint~\cite{HigWanBri19,JonEndMcA19}. 

In this work, we implement the VQD algorithm for excited-state simulations since VQD requires the same number of qubits as VQE and at most twice the circuit depth. In the VQD algorithm, the effective Hamiltonian is defined as~\cite{HigWanBri19}
\begin{equation}\label{eq:H_VQD} 
    \hat{H}_{\mathrm{VQD}} = \hat{H} + \sum_{I} \beta_I |\Psi_I\rangle \langle \Psi_I|,
\end{equation}
where $|\Psi_I\rangle$ is the previously found $I$-th eigenstate of $\hat{H}$ and $\beta_I$ is the weight of $I$-th eigenstate. Alternatively, we can also define the effective Hamiltonian with the projected operator
\begin{equation}
    \hat{P} = I - \sum_{I} |\Psi_I\rangle \langle \Psi_I|
\end{equation}
and 
\begin{equation}\label{eq:H_VQD2}
    \hat{H}_{\mathrm{VQD}} = \hat{P}^\dag\hat{H}\hat{P}.
\end{equation}
Eq.~\eqref{eq:H_VQD} and Eq.~\eqref{eq:H_VQD2} are equal when the previously found eigenstates are truly orthogonal. 



\section{Resource Estimation}\label{sec:resource}
The major challenge of simulating quantum chemistry on near-term noisy intermediate-scale hardware is the limited qubit coherence time, gate fidelities and quantum processors. The VQE algorithm is expected to be relatively robust to systematic errors from noise. At the same time, the VQE algorithm reduces the circuit depth at the cost a large number of measurements. Therefore, given the circuit depth is acceptable, the number of measurements is a crucial aspect to assess viability of a quantum algorithm.

Table~\ref{table:scaling} shows quantum resource requirement for UCCSD, UCCGSD, ADAPT, ADAPT-V and ADAPT-Vx. Here, we show the gate count for preparing the wave function and the number of measurements for performing the VQE algorithm. Note that we do not consider parallelization of implementing a quantum circuit, which depends on the architectural details of quantum hardware. UCCSD and UCCGSD considered as widely used wave function ansatzes are listed for comparison. As shown in Table~\ref{table:scaling}, the cost of energy measurement as quantified by terms in the Hamiltonian for UCCSD and UCCGSD simply scales as $\mathcal{O}((N-n)^2n^2)$ and $\mathcal{O}(N^4)$, respectively, for a gradient-free optimization algorithms and a fixed number of Trotter steps. Here, $n$ is the number of electrons. Meanwhile, the gate count for preparing both UCCSD and UCCGSD wave functions scales as $\mathcal{O}(N^4)$. On near-term quantum hardware, the circuit depth for state preparation rapidly become unaffordable as the size of system increases. 

The ADAPT generates a maximally compact wave function ansatz, which is particularly suitable for quantum hardware with limited coherence time. However, additional measurements for the residual gradients scale as $\mathcal{O}(N^8)$, which may be potentially efficiently computed with massively parallization in future. By contrast, ADAPT-V and ADAPT-Vx provide a good balance between the cost of state preparation and measurement. Analogous to UCCSD and UCCGSD, the cost of measurement for implementing the ADAPT-V and ADAPT-Vx VQE algorithm scales only as $\mathcal{O}(N^4)$ and $\mathcal{O}(N_mN^4)$, respectively. Meanwhile, their cost of state preparation scales as the size of parameters $\vec{\theta}$ in the wave function, which is often much less than $N^4$ (and $(N-n)^2n^2)$) as the size of system increases. Therefore, ADAPT-V and ADAPT-Vx present two promising wave function ansatzes for quantum simulations of practical chemical systems on near-term quantum hardware. 

Recently, several strategies have been proposed for measurement reduction in the framework of VQE\cite{McCRomBab16,VerYenYzm20,BonBabBri20,ZhaTraKir20,RalLovTra20}. Especially, Bonet-Monroig et al. has presented a scheme for sampling all elements of the fermionic 2-RDM using $\mathcal{O}(N^2)$ circuits, each of which requires only a local $\mathcal{O}(N)$-depth measurement circuit\cite{BonBabBri20}. Combining with these techniques, both the state preparation and measurement for ADAPT-V exhibits a significantly low circuit complexity, which improves the viability of quantum simulations of large systems. 

\begin{table}[!htb]
\centering \caption{Resource estimation for implementing the VQE algorithm with UCCSD, UCCGSD, ADAPT, ADAPT-RDM, ADAPT-V and ADAPT-Vx. $\times$ indicates that the residual gradients are necessary to be computed. $N$ is the number of spin orbitals and $n$ is the number of electrons. $N_s\approx N_uN_k$ is the size of parameters $\vec{\theta}$ in the wave function. $N_u$ is the number of operators to be updated at a time and $N_k$ is the number of iterations in different ADAPT ansates. $N_m$ is the number of operators in the auxiliary operator pool. ''Hamiltonian'' and ''Residual Gradients'' indicate the number of terms required to be measured in the Hamiltonian and the residual gradients.}
\label{table:scaling}
\begin{tabular}{|c|c|c|c|}
\hline
 & Gate count & Hamiltonian & Residual Gradients    \\
 \hline
UCCSD & $(N-n)^2n^2$ & $N^4$ & $\times$ \\
 \hline
UCCGSD & $N^4$ & $N^4$ & $\times$   \\
 \hline
ADAPT & $N_s$ & $N^4$ & $N^8$ \\
 \hline
 ADAPT-RDM & $N_s$ & $N^4$ & $N^6$ \\
 \hline
 ADAPT-V & $N_s$ & $N^4$ & $N^4$ \\
 \hline
 ADAPT-Vx & $N_s$ & $N^4$ & $N_mN^4$ \\
 \hline
\end{tabular}
\end{table}

\section{Results}\label{sec:results}

All calculations are performed with the modified ADAPT-VQE code~\cite{ADAPT-VQE}, which uses OpenFermion~\cite{openfermion} for mapping fermion operators onto qubit operators and PYSCF~\cite{pyscf} for all one- and two-electron integrals. The energy and wave function are optimized with the Broyden-Fletcher-Goldfarb-Shannon (BFGS) algorithm implemented in SciPy~\cite{scipy}. Gradients are computed with the analytical approach in Eq.~\eqref{eq:acse_grad}. Full configuration interaction results are used as the reference. The operator pool is composed of the spin-adapted operators in order to avoid the spin contamination. In this work, it is not able to use Eq.~\eqref{eq:acse_conv} as a convergence criteria for ADAPT-V and ADAPT-Vx since the approximate residual gradients are computed with Eq.~\eqref{eq:acse_grad_approx_1b} and Eq.~\eqref{eq:acse_grad_approx_2b}. Here, we adopt another convergence criteria using the covariance of the expectation of the Hamiltonian,
\begin{equation}\label{eq:conv_H}
     \delta H = \langle \Psi | \hat{H}^2 |\Psi\rangle - \langle \Psi | \hat{H} | \Psi \rangle ^2 < \epsilon
\end{equation}
Hereafter, for simplicity, we use ADAPT-V($N_u$) to indicate an ADAPT-V ansatz of Eq.~\eqref{eq:waveV} where $M$ contains $N_u$ operators and ADAPT-Vx($N_m$,$N_u$) to indicate an ADAPT-Vx ansatz where $M'$ contains $N_m$ operators and $M$ contains $N_u$ operators. ADAPT-Vx($N_u$) is an abbreviation of ADAPT-Vx($N_u$,$N_u$).

\begin{table*}[!htb]
\centering 
\caption{Errors of the energy (in mHartree), the size of parameters $\vec{\theta}$ ($N_{s}$) and the number of iterations ($N_k$) in the ADAPT, ADAPT-V and ADAPT-Vx with different $N_u$. $N_u$ is the number of operators to be updated at a time. $N_m$ is the number of operators in the auxiliary operator pool $M'$ ( $N_m=30$ if $N_u<30$, otherwise $N_m=N_u$ ).}\label{table:conv}
\begin{tabular}{cccccccccccccccc}
\hline \hline
&\multicolumn{3}{c}{ADAPT-V($N_u$}) && \multicolumn{3}{c}{ADAPT-Vx($N_u$)} && \multicolumn{3}{c}{ADAPT-Vx($N_m$,$N_u$)} &&\multicolumn{3}{c}{ADAPT}\\
\cline{2-4}\cline{6-8} \cline{10-12}\cline{14-16}
$N_u$ & Energy & $N_{s}$ & $N_{k}$ && Energy & $N_{s}$ & $N_{k}$&& Energy & $N_{s}$ & $N_{k}$ && Energy & $N_{s}$ & $N_{k}$\\
\hline
1 & 3.08E+01 & 10  & 10 && 3.08E+01 & 10  & 10 && 4.72E-02 & 74  & 74 && 5.88E-02 & 63  & 63 \\
5 & 3.96E+00 & 65  & 13 && 4.04E+00 & 85  & 17 && 4.17E-02 & 80  & 16 && 4.03E-02 & 80  & 16 \\
10& 3.94E+00 & 70  & 7  && 3.95E+00 & 60  & 6  && 5.41E-02 & 80  & 8  && 3.96E-02 & 80  & 8  \\
20& 6.42E-02 & 120 & 6  && 7.45E-02 & 100 & 5  && 2.74E-02 & 100 & 5  && 2.49E-02 & 100 & 5  \\
30& 4.55E-02 & 115 & 4  && 2.45E-02 & 115 & 4  && 2.45E-02 & 115 & 4  && 1.80E-02 & 115 & 4  \\
40& 4.17E-02 & 105 & 3  && 4.95E-02 & 105 & 3  && 4.95E-02 & 105 & 3  && 2.09E-02 & 105 & 3  \\
50& 1.45E-02 & 125 & 3  && 1.41E-02 & 125 & 3  && 1.41E-02 & 125 & 3  && 9.74E-03 & 125 & 3  \\
\hline \hline
\end{tabular}
\end{table*}

\subsection{$H_6$ molecule}
Hydrogen chain is an interesting model system to explore elemental physical phenomena in modern condensed matter physics, such as an antiferromagnetic Mott phase and an insulator-to-metal transition~\cite{WelRuszgi16,Ruszgi16,MarClaFen19,LiuSheZha20}. Here, a $H_6$ molecule with each hydrogen atom equispaced along a line is used to assess different variational approaches.

Firstly, we test the performance of ADAPT, ADAPT-V and ADAPT-Vx with different number of operators to be updated in each iteration. Table~\ref{table:conv} shows errors of the energy, the total number of parameters and iterations as the function of $N_u$ for $H_6$ at $R=1.5$ \r{A}. The convergence thresh $\epsilon$ is set to be $1\times 10^{-4}$. ADAPT-V and ADAPT-Vx VQE algorithms stop if either the convergence criteria is satisfied or the energy change is less than $1\times 10^{-10}$ Hartree. 

As expected, the converged energy is not significantly affected by $N_u$ since the accurate residual gradients of Eq.~\eqref{eq:acse_res} are computed in ADAPT. No matter how many operators are updated in each iteration, the ordering of operators is always determined according to corresponding magnitude of the residual gradients, which guarantees that the greatest amount of correlation effect is restored. For different $N_u$ values, the ADAPT energy converges to a higher precision at the cost of a longer operator sequence. For example, the number of parameters in ADAPT(50) is $\sim 2$ times of that in ADAPT(1) while the error of the energy slightly decreases from $5.88\times 10^{-2}$ to $9.74\times 10^{-3}$ mHartree. In addition, as $N_u$ increases, the number of iterations significantly decreases so that the total number of energy evaluations will be also significantly reduced. Therefore, it is important to choose an appropriate $N_u$ in order to achieve a delicate balance between the gate count and measurement. 

ADAPT-V($N_u$) and ADAPT-Vx($N_u$) exhibit very similar errors of the energy. For a small $N_u$ value, both ADAPT-V($N_u$) and ADAPT-Vx($N_u$) are hard to achieve the chemical accuracy due to approximate residual gradients. Especially, the energy computed with ADAPT-V(1) or ADAPT-Vx(1) significantly deviates from the exact energy with the error of 30.8 mHartree. ADAPT-V(5) and ADAPT-Vx(5) perform much better but the corresponding VQE procedures still stop after 13 and 17 iterations, respectively, because the energy change is too small. This reveals that it is not able to improve the convergence by only adjusting the ordering of the operator sequence. As $N_u$ increases, errors of the energy for ADAPT-V($N_u$) and ADAPT-Vx($N_u$) significantly decreases. Therefore, operators that are identified as important can be included in the operator sequence as long as the $N_u$ value is large enough.  

As shown in ADAPT-V($N_u$) and ADAPT-Vx($N_u$), $N_u=30$ is large enough to converge the energy. Hence, in ADAPT-Vx($N_m$,$N_u$), the number of operators, $N_m$, in the auxiliary operator pool is set to be greater than or equal to 30. When $N_u > 30$, $N_m$ is equal to $N_u$. The overall performance of ADAPT-Vx($N_m,N_u$) is very close to that of ADAPT. Especially, the size of the operator sequence and the number of iterations are almost the same in these two ansatzes. This indicates that ADAPT-Vx($N_m,N_u$) is able to generate an ansatz as maximally compact as ADAPT while the number of measurements required to compute residual gradients is significantly reduced.

In principle, ADAPT with a small $N_u$ generates a low depth circuit ansatz while ADAPT-V and ADAPT-Vx require a larger $N_u$ which means a more complex circuit to achieve the same accuracy. However, on noisy quantum hardware, errors for computing residual gradients are unavoidable due to imperfect fidelity and measurement. Therefore, a large $N_u$ is often necessary to implement ADAPT on near-term quantum hardware. As $N_u$ is greater than or equal to 20, the performance of ADAPT-V($N_u$) and ADAPT-Vx($N_u$) is comparable to ADAPT. Considering errors for measuring the 2-RDM on quantum hardware, ADAPT-V($N_u$) and ADAPT-Vx($N_u$) may still require a larger $N_u$ value than ADAPT. However, the difference between ADAPT-V (or ADAPT-Vx) and ADAPT is expected to be quite small when they are implemented on noisy quantum hardware.

\begin{figure}[!htb]
\begin{center}
\includegraphics[width=0.45\textwidth]{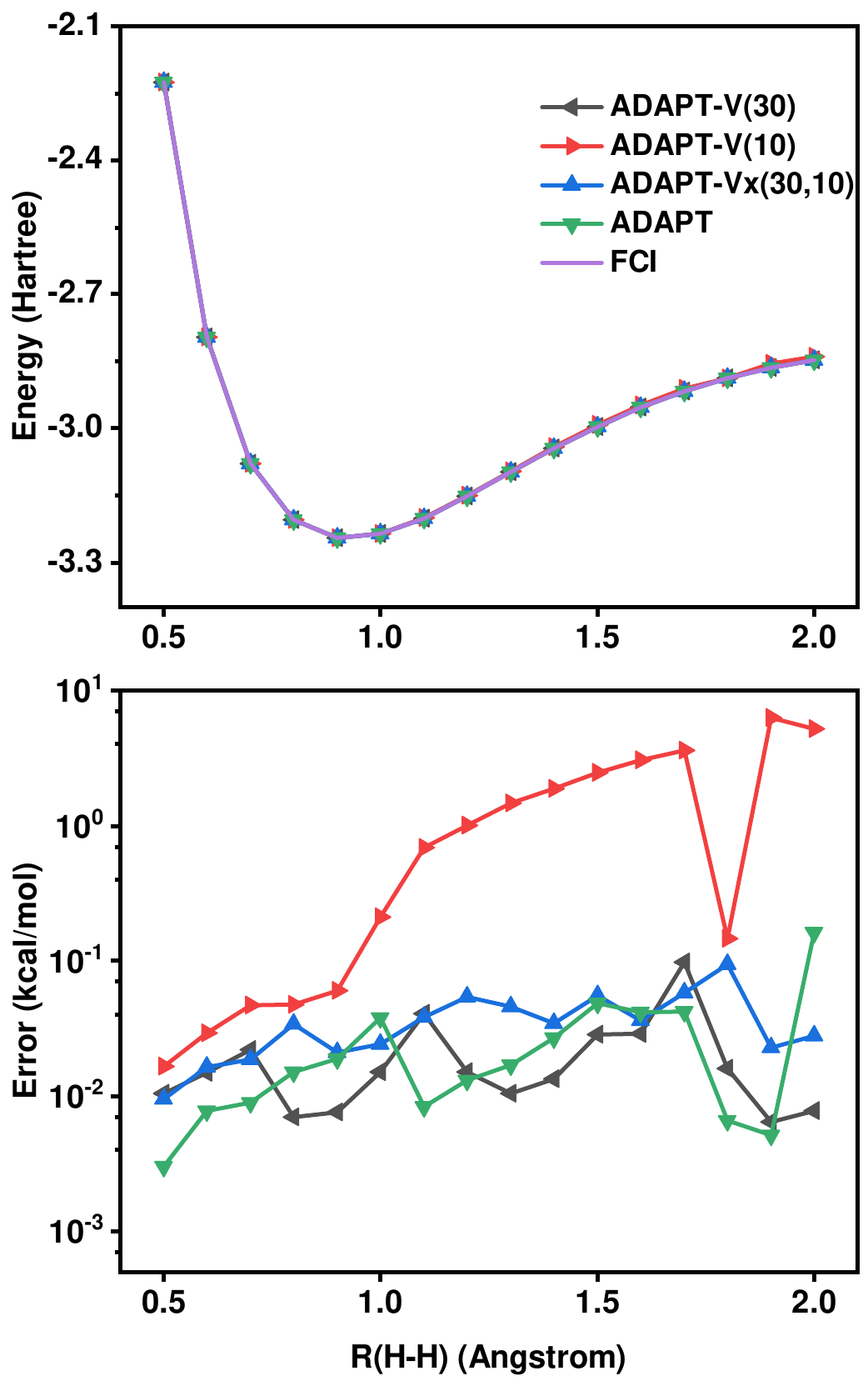}
\end{center}
\caption{The ground-state potential energy curve and absolute energy error
with respect to the FCI result for $H_6$ computed
with ADAPT-V(30), ADAPT-V(10), ADAPT-Vx(30,10) and ADAPT.} \label{fig:H6S0}
\end{figure}

In Figure~\ref{fig:H6S0}, we present the ground-state potential energy curve and the absolute energy error with respect to the FCI results for ADAPT-V(10), ADAPT-V(30), ADAPT-Vx(30,10) and ADAPT. Here, $\epsilon$ is set to be $1\times 10^{-4}$. In previous studies, ADAPT had been demonstrated to be quite accurate for small molecules, such as LiH, BeH$_2$ and H$_6$. ADAPT is expected to be able to reproduce the FCI result since the exact wave function can be well approximated as a series of two-body unitary transformation acting on a reference wave function. For the ground state of $H_6$ molecule, the maximum deviation of ADAPT with respect to FCI is 0.16 kcal/mol at 2.0 \r{A}. 

ADAPT-V(30) produces almost exactly the same potential energy curve as ADAPT. As R(H-H) increases from 0.5 to 2.0 \r{A}, ADAPT-V(30) shows a random fluctuation of errors in energy that are one order of magnitude less than 1 kcal/mol. This means that for $H_6$ molecule $N_u=30$ is large enough to suppress the influence of approximate residual gradients computed with the cumulant 3-RDM. This conclusion is consistent with the analysis as discussed above. In contrast, ADAPT(10) shows an overall monotonic increase of errors as the bond length of Hydrogen-Hydrogen elongates, that is, the correlation effect becomes stronger. At 1.9 \r{A}, ADAPT(10) exhibits a maximum deviation of 6.33 kcal/mol in energy. Therefore, for the strongly correlated systems, the Valdemoro's reconstruction of 3-RDM from the 2-RDM is not able to give an accurate description of the importance of operators. In addition, there is a sudden decreasing of the error when the bond length increases from 1.8 \r{A} to 1.9 \r{A}. This discontinue mainly results from the small $N_u$ for the scanning of the potential surface if some significant operators are occasionally included for one structure while missed for another structure.

The size of the auxiliary operator pool $M'$ for ADAPT-Vx(30,10) is the same as the number of updated operators for ADAPT-V(30). Therefore, a very similar performance of ADAPT-V(30) and ADAPT-Vx(30,10) on the simulation of the potential energy surface for $H_6$ molecule is foreseeable. As shown in Table~\ref{table:H6}, ADAPT-V(30) and ADAPT-Vx(30,10) exhibit the same NPE of 0.09 kcal/mol. In addition, although in each iteration 10 operators have been updated in both ADAPT-Vx(30,10) and ADAPT-V(10), ADAPT-Vx(30,10) is much more stable and accurate than ADAPT-V(10) as R(H-H) increases. It is clear that 10 operators refined from the auxiliary operator pool $M'$ with exact residual gradients are much better than those 10 operators with largest gradients in $M'$.

Figure~\ref{fig:H6S0} shows the first excited-state potential energy curve and the absolute energy errors with ADAPT-V(10), ADAPT-V(30), ADAPT-Vx(30,10) and ADAPT. The quality of the ground state is crucial for excited-state calculations within the VQD framework. Here, all ground state wave functions are prepared with ADAPT since it has been demonstrated to be accurate enough for ground state calculations. In addition, because spin-adapted operators have been used in this work to avoid the spin contamination, it is necessary to perform VQD simulations with different reference states in order to optimize the wave function to the target state. For the $H_6$ molecule, both HOMO$\rightarrow$LUMO and HOMO$\rightarrow$LUMO+1 excitation configurations have been used as the reference state, respectively, and the optimized wave function with a lower excitation energy is considered as the first excited state.

Unlike the ground state case, all of four wave function ansatzes yield quite promising results with a large error of 0.23 kcal/mol for ADAPT-V(10). It is worth mentioning that as the orthogonal constraint condition is introduced, the converged excited-state energies is possibly slightly lower than corresponding FCI energies. For example, the first excited-state energies computed with ADAPT(30) and FCI are -2.8221455 and -2.8221449 Hartree, respectively. This mainly results from the numerical precision of the ground state wave function. Therefore, it is able to avoid this problem with the ground state obtained from ADAPT using a tighter convergence criteria.  

In Table~\ref{table:H6}, we present the nonparallelity error (NPE) for the ground state and the first excited state simulations with different ADAPT ansatzes. NPE is defined as the difference between the maximum and minimum error, which is a useful measure of performance, since we are interested in relative energetics in most chemical applications. In the ground state and first excited states, ADAPT-V(10) is the worst one in four wave function ansatzes in terms of the NPE.

\begin{figure}[!htb]
\begin{center}
\includegraphics[width=0.45\textwidth]{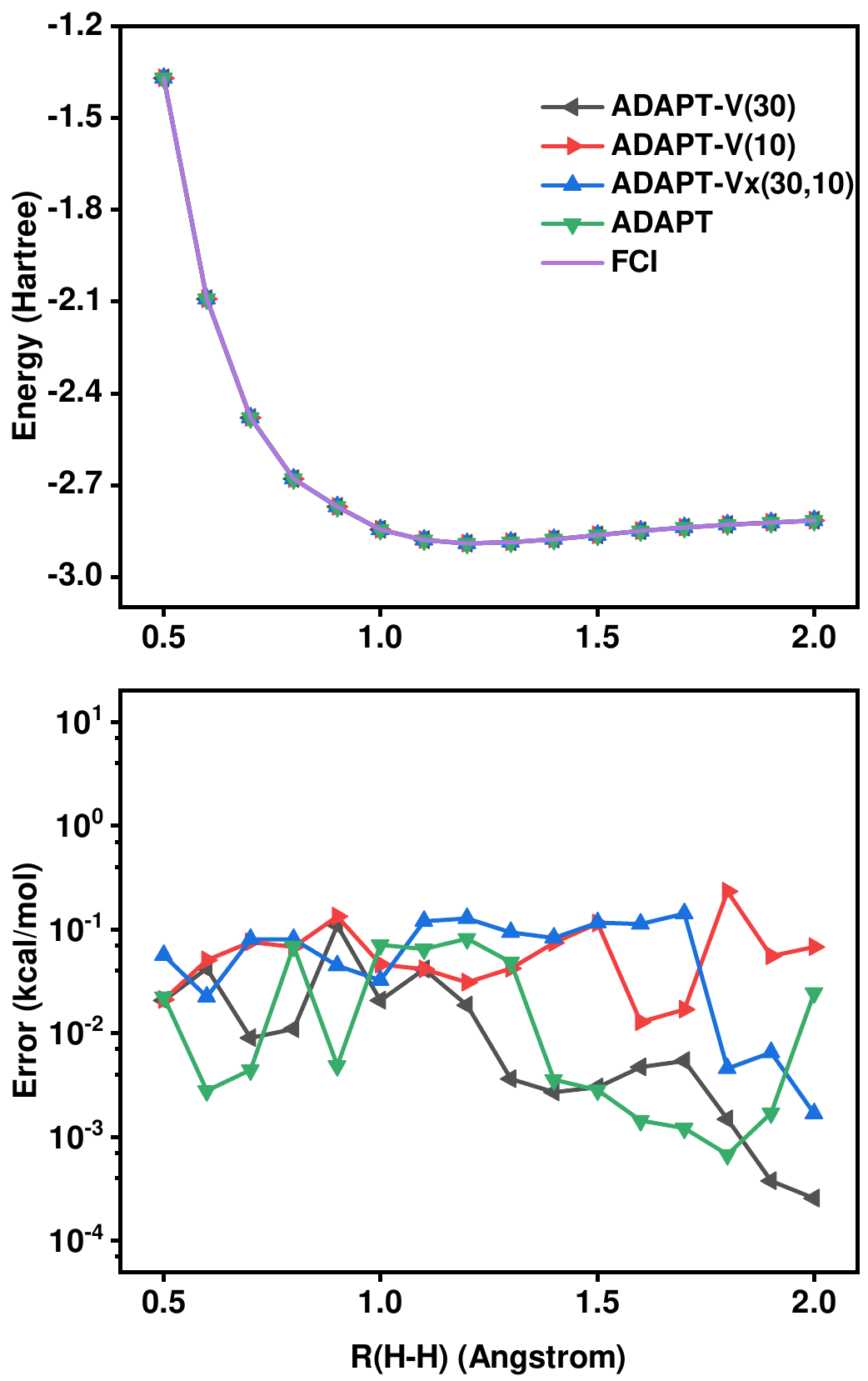}
\end{center}
\caption{The first excited-state potential energy curve and absolute energy error
with respect to the FCI result for $H_6$ computed
with ADAPT-V(30), ADAPT-V(10), ADAPT-Vx(30,10) and ADAPT.} \label{fig:H6S1}
\end{figure}

\begin{table}[!htb]
\centering \caption{Nonparallelity error (NPE) (in kcal/mol) in the ground state ($S_0$) and first excited state ($S_1$) of $H_6$ for various variational approaches. }
\label{table:H6}
\begin{tabular}{|c|c|c|}
\hline
&$S_0$&$S_1$\\
\hline
ADAPT-V(30) &0.09&0.11\\
\hline
ADAPT-V(10) & 6.31&0.22\\
\hline
ADAPT-Vx(30,10) & 0.09 & 0.14\\
\hline 
ADAPT & 0.16 & 0.08 \\
\hline
\end{tabular}
\end{table}

\subsection{$N_2$ molecule}
The dissociation of $N_2$ is very challenging for many traditional electronic structure methods as it involves the breaking of a triple bond~\cite{Sie83,MaLiLi06}. At stretched bond lengths, there are a total of 6 electrons that are strongly entangled. Therefore, UCC with single and double excitations is hard to correctly describe the dissociation potential energy surface. It often should include triples, quadruples and higher excitations to obtain an accurate result. The ADAPT ansatz is expected to provide a correct description of the strongly correlated effect. Here, we apply ADAPT to study the dissociation of $N_2$. All calculation are performed with $1s^22s^2$ electrons frozen, which result in a model system that contains 6 electrons and 6 orbital. Here, we set $\epsilon=1\times 10^{-4}$ for the ground state and $\epsilon=2\times 10^{-5}$ for the first excited state ($S_1$) since the accurate description of $S_1$ of $N_2$ is extremely challenging. 

\begin{figure}[!htb]
\begin{center}
\includegraphics[width=0.45\textwidth]{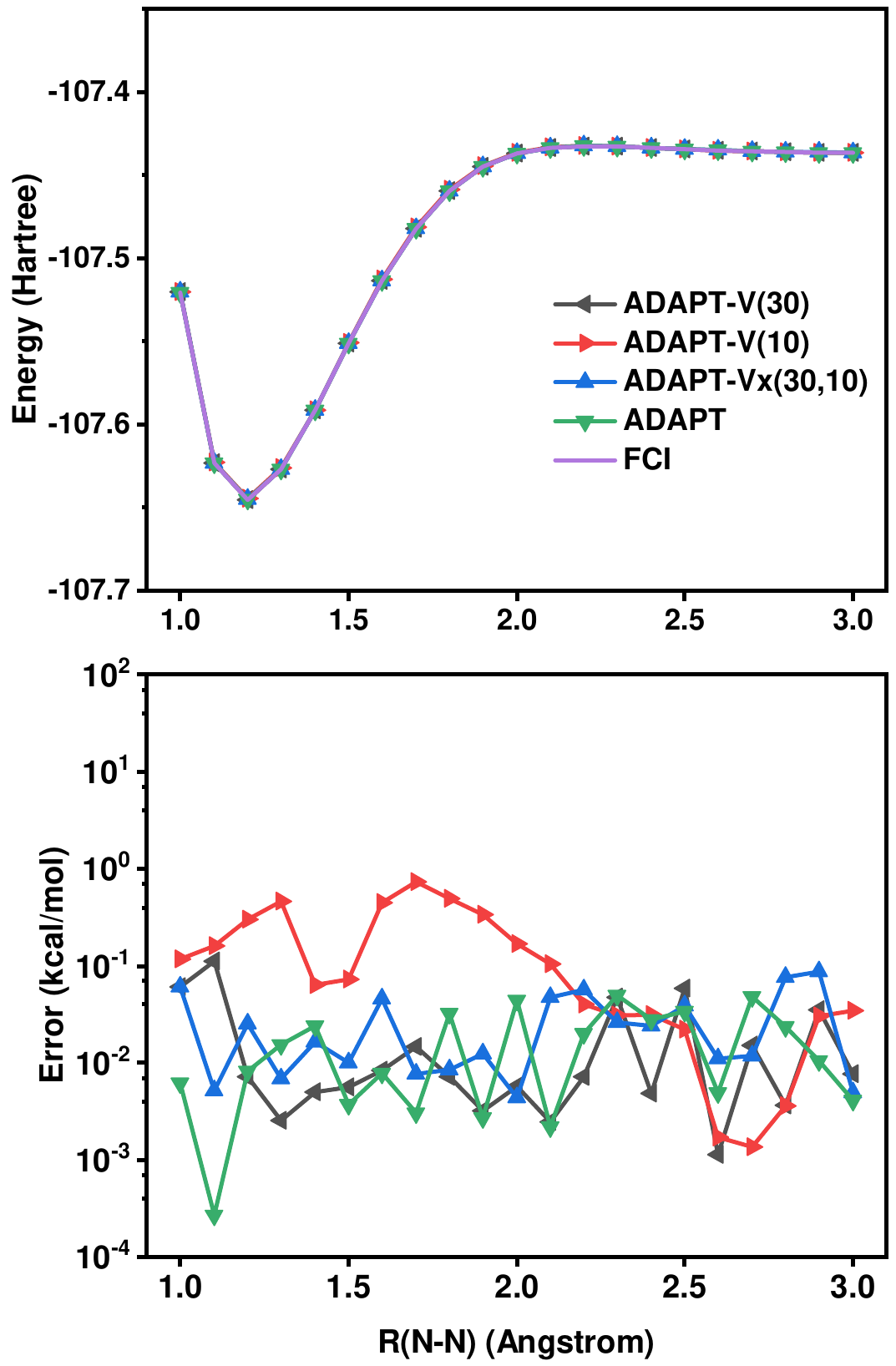}
\end{center}
\caption{The ground-state potential energy curve and absolute energy error
with respect to the FCI result for $N_2$ computed
with ADAPT-V(30), ADAPT-V(10), ADAPT-Vx(30,10) and ADAPT.} \label{fig:N2S0}
\end{figure}

Figure~\ref{fig:N2S0} shows the ground-state potential energy curve and the absolute energy error as a function of N-N bond length for ADAPT-V(10), ADAPT-V(30), ADAPT-Vx(30,10) and ADAPT. The overall errors of ADAPT-V(30), ADAPT-Vx(30,10) and ADAPT are less than 0.1 kcal/mol except for the error of ADAPT-V(30) at 1.1 \r{A}, which is 0.11 kcal/mol. Unlike the ground state simulation of $H_6$, the ground-state dissociation curve computed with ADAPT-V(10) agrees very well with the FCI curve. However, the NPE of ADAPT-V(10) is 0.73 kcal/mol, which is much larger than other three ansatzes. 

As the Nitrogen-Nitrogen bond length increases, the character of the first excited state of $N_2$ is keeping changing. Here, four reference states including the Hartree-Fock ground state and three single excitation configurations (HOMO$\rightarrow$LUMO, HOMO$\rightarrow$LUMO+1, HOMO-2$\rightarrow$LUMO) have been used to obtain an accurate description of the first excited state. The overall NPEs for $S_1$ of $N_2$ are much larger than those for the ground state. Analogous to the performance of ADAPT-RDM(10) for the ground state simulation of $H_6$, the NPE of ADAPT-RDM(10) is as large as 4.38 kcal/mol for $S_1$ of $N_2$. In addition, there is also a sudden increasing of the error of the energy when the bond length increases from 1.9 \r{A} to 2.0 \r{A}. When the N-N bond elongates, especially larger than 2.4 \r{A}, all of four ansatzes exhibit relatively large deviations of the energy because there exist a larger number of near-degenerate excited states.

\begin{figure}[!htb]
\begin{center}
\includegraphics[width=0.45\textwidth]{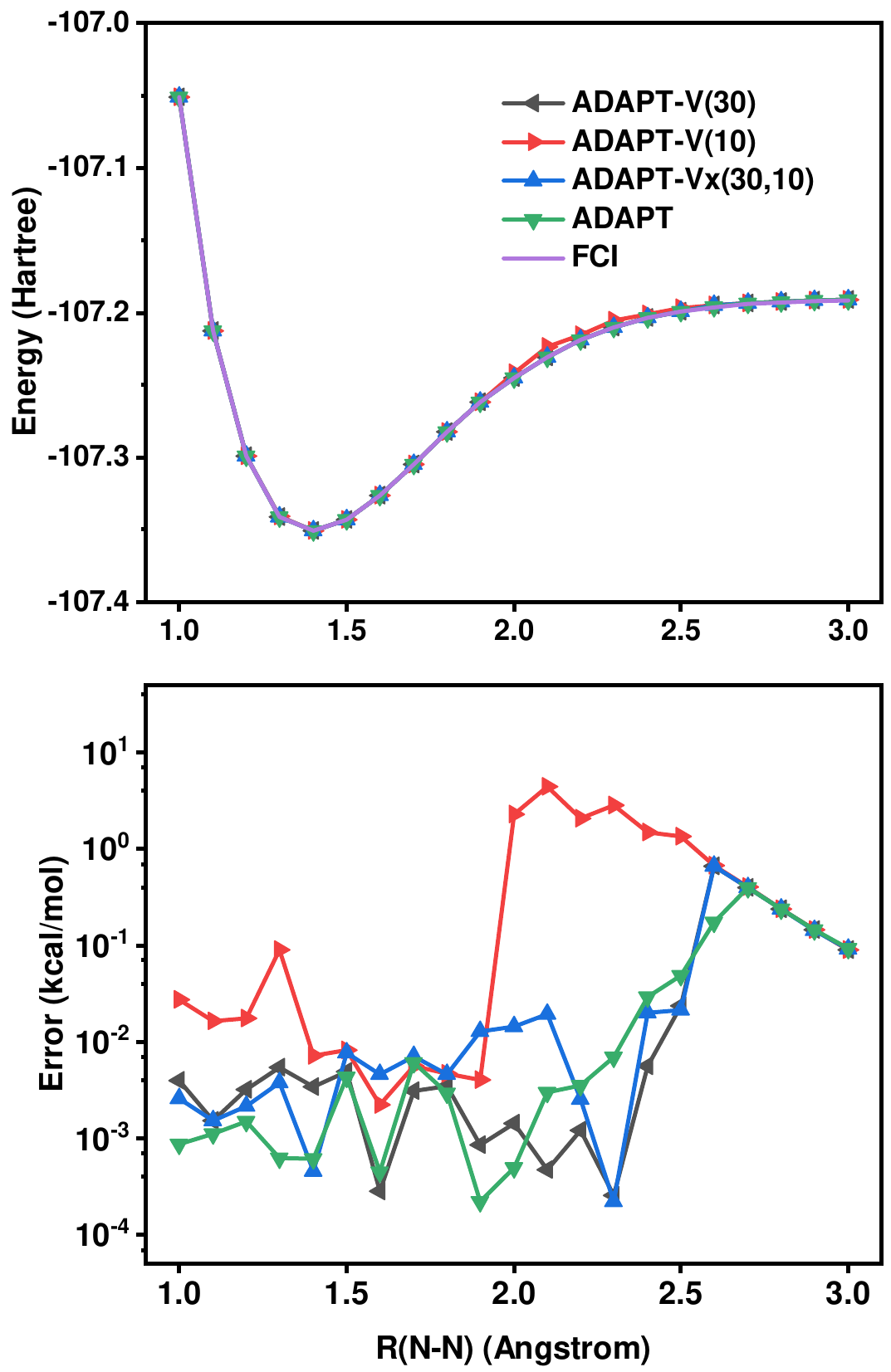}
\end{center}
\caption{The first excited-state potential energy curve and absolute energy error
with respect to the FCI result for $N_2$ computed
with ADAPT-V(30), ADAPT-V(10), ADAPT-Vx(30,10) and ADAPT.} \label{fig:N2S1}
\end{figure}

\begin{table}[!htb]
\centering \caption{Nonparallelity error (NPE) (in kcal/mol) in the ground state ($S_0$) and first excited state ($S_1$) of $N_2$ for various variational approaches.}
\label{table:N2}
\begin{tabular}{|c|c|c|}
\hline
&$S_0$&$S_1$\\
\hline
ADAPT-V(30) &0.11&0.66\\
\hline
ADAPT-V(10) &0.73&4.38\\
\hline
ADAPT-Vx(30,10) & 0.08 & 0.66\\
\hline 
ADAPT & 0.05 & 0.40 \\
\hline
\end{tabular}
\end{table}

\section{Conclusion}\label{sec:conclusion}

In this work, we present an efficient adaptive variational quantum algorithm based on density matrix matrices. The original ADAPT ansatz generate a maximally compact operator sequence at the cost of a huge number of measurements that scales as $\mathcal{O}(N^8)$. In order to reduce this overhead, we firstly reformulate the residual gradients using the 2-RDM and 3-RDM, which reduces the scaling of measurement to $\mathcal{O}(N^6)$. Furthermore, we propose to build the 3-RDM as approximate functionals of the 2-RDM using Valdemoro's reconstruction. We refer this ansatz as ADAPT-V, which requires only the measurement of 2-RDM on quantum hardware to determine the optimal operator sequence. Therefore, the number of measurements for computing the energy and gradients in ADAPT-V scales only as $\mathcal{O}(N^4)$. In addition, we introduce an ADAPT-Vx ansatz that refines operators to be updated with exact residual gradients from a large auxiliary operator pool. The ADAPT-Vx ansatz is more stable than the ADAPT-V ansatz with the number of update operators being small enough while at a slightly increasing measurements. 

The performance of ADAPT-V and ADAPT-Vx has been assessed by computing the ground-state and exicted state potential energy curve of two strongly correlated systems, a finite Hydrogen chain $H_6$ and a triple-bond molecule $N_2$. The benchmark calculations reveal that ADAPT-V and ADAPT-Vx with moderate $N_u$ are accurate for both ground and excited states. Considering the reduced circuit depth and measurement, the performance of ADAPT-V and ADAPT-Vx is particularly encouraging for simulating chemical systems on near-term quantum computers. 

\section{Acknowledgments}
This work is supported by the National Natural Science Foundation of China (21688102, 21825302, 22073086, 21803065), by National Key Research and Development Program of China (2016YFA0200604), Anhui Initiative in Quantum Information Technologies (AHY090400), the Fundamental Research Funds for the Central Universities (WK2060000018).

\footnotesize{

}

\vspace{3ex}

\end{document}